\documentclass[aps,groupedaddress,twocolumn,twoside,showpacs,final]{revtex4}
\usepackage{amsmath}
\usepackage{amssymb}
\usepackage{amsfonts}
\usepackage{epsfig}
\usepackage{times}

\def\bs{\backslash}
\def\emb{\hookrightarrow}
\newcommand\abs[1]{| #1 |}
\newcommand\map[3]{#1 : #2 \rightarrow #3}
\newcommand\imap[3]{#1 : #2 \mapsto #3}
\newcommand\mapi[2]{#1 \mapsto #2}
\def\polylog{\operatorname{polylog}}
\def\lr{\leftrightarrow}
\def\idty{\openone}

\newcommand\set[1]{\{ #1 \}}
\newcommand\setcond[2]{\{ #1 \,|\, #2 \}}

\def\cG{{\mathcal G}}
\def\cH{{\mathcal H}}
\def\cK{{\mathcal K}}
\def\bR{{\mathbf R}}
\def\bT{{\mathbf T}}
\def\P{{\mathbb P}}
\def\Z{{\mathbb Z}}

\def\etal{{\em et al.}}

\begin{document}
\title{Scaling and renormalization in fault-tolerant quantum
  computers\footnote{Based on the talk given at the Simons Conference
    on Quantum and Reversible Computation, Stony Brook NY, May 28-31, 2003}}

\author{Maxim Raginsky}\email{maxim@ece.northwestern.edu}
\affiliation{{\it Center for Photonic Communication and Computing,
    Department of Electrical and Computer Engineering, Northwestern
    University, Evanston, IL 60208, USA}}

\begin{abstract}
This work is concerned with phrasing the concepts of
fault-tolerant quantum computation within the framework of disordered
systems, Bernoulli site percolation in particular. We show how the
so-called "threshold theorems" on the possibility of fault-tolerant
quantum computation with constant error rate can be cast as a
renormalization (coarse-graining) of the site percolation
process describing the occurrence of errors during computation. We
also use percolation techniques to derive a trade-off between the
complexity overhead of the fault-tolerant circuit and the threshold error rate.

\end{abstract}
\maketitle

Many researchers' confidence in the eventual
experimental realization of reliable large-scale quantum computation
rests upon a number of results known collectively as
the ``threshold theorem'' (see, e.g., \cite{kit,klz,preskill,ab}). The
recurring motif of these theorems is that, under certain reasonable
assumptions on errors in the computer, fault-tolerant
quantum computation is possible provided that the error rate $\eta$ does not
exceed some {\em threshold value} $\eta_c$.

Among the papers dealing with threshold theorems, the work of Aharonov
and Ben-Or \cite{ab} is especially noteworthy: it contains rigorous
constructive proofs of the possibility of arbitrarily reliable
sub-threshold quantum computation for a wide variety of noise models,
including local probabilistic noise (i.e., when each gate in the
computer suffers an error with probability $\eta$ and functions
correctly with probability $1-\eta$, independently of all other gates both in space and in time),
noise with exponentially decaying space-time correlations, and general
noise (i.e., one not described by an {\em a priori} probabilistic
model). The threshold theorem is also specialized to quantum computation on the
$d$-dimensional hypercubic lattice $\Z^d$ with a restriction that
only nearest-neighbor qubits can interact directly.

The strategy of Aharonov and Ben-Or has a lot in common with the work
of G\'acs \cite{gac} on fault-tolerant classical
computation in cellular automata. In particular, at the core of their
proof lies the idea of iterated simulation of one unreliable computer
by another, with (quantum) error correction implemented on all levels
of iteration. It is then shown that, provided that the error rate
$\eta$ (say, per quantum gate) is below a certain threshold $\eta_c$,
each iteration will reduce the {\em effective} error rate (rate at
which errors occur in the encoded information). One of the goals of this paper
is to provide an alternative interpretation of the Aharonov--Ben-Or
method for local probabilistic noise in terms of a simple disorder
model, namely Bernoulli site percolation \cite{hug}. In particular, we
will exhibit a close relation between the recursive simulation
technique and a renormalization (coarse-graining) of the site
percolation process, and then use this relation to derive a trade-off
between the threshold error rate $\eta_c$ and the complexity-theoretic
overhead required to implement the computation fault-tolerantly, i.e.,
the minimum number of iterations needed to bring the effective error
rate down to some desired level.

Let us
sketch very briefly the key ideas behind iterated simulation
\cite{ab}. The basic ingredient is a quantum error-correcting code
\cite{kl}, i.e., an isometric embedding of a Hilbert space $\cH_k$ of
$k$ qubits as a $k$-qubit subspace $\cK$ of some Hilbert space $\cH_m$ of
$m$ qubits, with $m \ge k$ [this is referred to as a quantum $(m,k)$-code]. Aharonov and Ben-Or \cite{ab} use quantum $(m,1)$-codes. Errors
are modeled by {\em quantum operations} \cite{kl}, i.e., mappings of
the form $\bT(\rho) = \sum_j E_j \rho E^\dag_j$, where $\rho$ is a state (density matrix) on $\cH_m$, and the
operators $\map{E_j}{\cH_m}{\cH_m}$ (called the {\em Kraus operators} of $\bT$) satisfy the constraint $\sum_j
E^\dag_j E_j = \idty$. A quantum $(m,1)$-code is said to {\em
  correct $d$ errors} (with $d \le m$) if there exists a quantum operation
$\bR$ (the {\em recovery operation}), such that for any density matrix supported on the code space
$\cK$ and for any operation $\bT$ whose Kraus operators are tensor
products of at most $d$ nontrivial single-qubit operators acting on
the qubit components of $\cK$ and identity
operators on the rest of the qubits, we have $\bR\circ \bT(\rho) =
\rho$. If this is the case, we say that the code $\cH_k \emb \cK
\subseteq \cH_m$ is a
quantum $(m,k,d)$-code.

When quantum error-correcting codes are used to protect information
inside a quantum computer, each qubit is encoded separately. Thus, if
we use a quantum $(m,1)$-code and there are $n$ qubits, then the
encoded state of the computer's register is a state of $O(mn)$ qubits (the
constant hidden in the ``big-oh'' notation reflects the ancillae added
to each $m$-qubit block in order to implement the recovery operations). We
assume that all encoding, decoding, recovery, and computation
operations are implemented using quantum gates from a suitable universal set $\cG$
\cite{bar}. Aharonov and Ben-Or use {\em concatenated codes}. That is,
each qubit is encoded in a block of $m$ qubits, each of
which is in turn encoded in a block of $m$ qubits, and so on. If we do
$r$ levels of concatenation, we end up with $n$ blocks of $m^r$ qubits
each. At the end of computation decoding proceeds
hierarchically, starting with the highest (coarse-grained) level and
ending with the lowest (fine-grained) level.

The computation itself is also defined hierarchically. Consider first the case of $r=1$ (one level of
concatenation). Each gate $U \in \cG$ in the original circuit now
corresponds to a certain composition of gates from $\cG$, referred to as the {\em procedure of $U$}. Each time step of the original (unencoded)
computation now corresponds to a {\em working period}, consisting of
two stages: the recovery operation applied on each $m$-qubit
block, followed by parallel application of the procedures of all the
gates used in the original computation in this particular time
step. Denoting the original quantum circuit by $C_0$ and the encoded
circuit by $C_1$, we have a mapping $\imap{M}{C_0}{C_1}$. This is,
essentially, a {\em simulation} of the original circuit $C_0$ by the
encoded circuit $C_1$.

Now imagine the same construction done with the circuit $C_1$,
treating each $m$-qubit block as a unit (a $1$-block, in the
terminology of Aharonov and Ben-Or \cite{ab}), and each procedure as a single gate. This gives a circuit $C_2 = M(M(C_0))
\equiv M^2(C_0)$. After $r$ levels of concatenation we will end up
with a circuit $C_r = M^r(C_0)$ that simulates the circuit $C_{r-1}$ that
simulates the circuit $C_{r-2}$, and so on. The coarse-grained
computation in $C_r$ takes place on $n$ $r$-blocks, each of which
consists of $m^r$ qubits on the fine-grained level. Again, at the end of
computation decoding takes place recursively, starting at the level
of $r$-blocks and so on, until the level of individual qubits
($0$-blocks) is reached.

Note that the concatenated circuits thus constructed are
essentially {\em self-similar}. That is, if we start with
$C_{r-s}$ and do $s$ levels of concatenation, then the resulting
circuit $C_r$ will ``look like'' the circuit $C_s \equiv M^s(C_0)$
if each block of $m^s$ qubits in $C_r$ is treated as a unit (an $s$-block
\cite{ab}).

Finally, we have to safeguard the computation against the propagation of
errors. The idea is to design all encoding/decoding operations and all
procedures in such a way that they are {\em
  fault-tolerant}. Informally, this means that a small number of
localized errors during a procedure will not affect too many
qubits at the end of the procedure, so that the recovery operation
which is applied during the next working period will be successful
with high probability \cite{ab}. More precisely, given a quantum
$(m,1,d)$-code, we say that it has {\em spread} $s$ if
a single error anywhere during a procedure results in at
most $s$ faulty qubits in each block at the end of the procedure. Note
that it follows from self-similarity that this
definition of the spread of the code makese sense at any level in
the concatenation hierarchy. Now if the code under scrutiny is a
quantum $(m,1,d)$-code, then we require that $s \le d$, so that at
least one error can be tolerated in each procedure. Aharonov and
Ben-Or \cite{ab} call such codes {\em quantum computation codes}.

\begin{figure}
\includegraphics[width=\columnwidth]{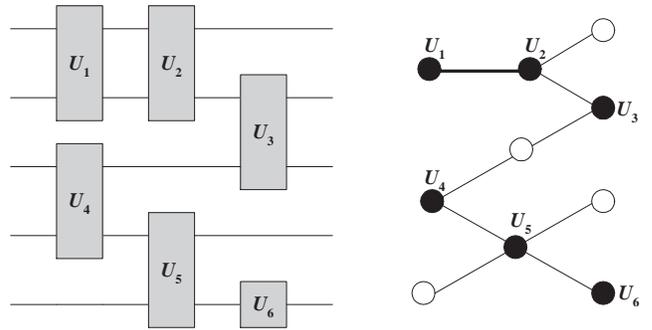}
\caption{A simple quantum circuit and the corresponding interaction
  graph. The solid vertices correspond to nontrivial quantum gates,
  the empty vertices correspond to identity gates. Thick edges are
  those that resulted from the collapsing operation on the
  intermediate multigraph.}
\label{fig:ckt}
\end{figure}

In
order to visualize the occurrence of errors during computation, we
will associate an {\it interaction graph} to the quantum circuit under
consideration. Given a circuit $C$, we will construct its interaction
graph $\Gamma_C$ in two steps. First, we will replace each quantum
gate in $C$, including the identity gates, by a single vertex. This
intermediate combinatorial object will in general be a {\it multigraph},
i.e., there can be multiple edges joining a given pair of
vertices. The second step is to collapse each such bundle of
multiple edges into a single edge. This construction is illustrated
for a simple quantum circuit in Fig.~\ref{fig:ckt}. In precise terms,
the interaction graph corresponding to the quantum circuit
$C$ is given by $\Gamma_C = (V_C,E_C)$, where
the vertex set $V_C$ is the set of all quantum gates, including the
identity gates, that are used in the computation, and the edge set
$E_C$ consists of pairs $(v,v')$, $v,v' \in V_C$, such that there is
at least one quantum wire in $C$ connecting the gates corresponding to
$v$ and $v'$. We will use the shorthand $v \lr v'$ to denote the fact
that $(v,v') \in E_C$.

The vertices of $\Gamma_C$ can now be thought of as the potential
locations of errors, so that an error that occurs at some $v \in V_C$
can propagate to some or all of $v' \in V_C$ for which $v \lr
v'$. Assuming that there are no correlations between errors in distinct
gates, either in space or in time (which is the focus of this paper), we lose no generality if we suppose that $\Gamma_C$ is
{\it connected}, i.e., there exists a path between any two $v,v' \in
\Gamma_C$. (Otherwise we could represent $\Gamma_C$ as a union of
maximal connected subgraphs, $\Gamma_C = \Gamma^1_C \cup \ldots \cup
\Gamma^L_C$, so that if an error occurs in some $\Gamma^i_C$, $1\le i
\le L$, it will not propagate to any $\Gamma^j_C$ with $j \neq i$.)
Another important feature of the graph $\Gamma_C$ is that, even if it
corresponds to an ``infinitely large'' quantum circuit, its vertices have
{\it bounded coordination number}, i.e.,
$$
\max_{v \in V_C} \#
\left(\setcond{v' \in V_C}{v \lr v'}\right) < \infty,
$$
where $\#(S)$
denotes cardinality of the set $S$. Basically, this is a consequence of
the fact that all quantum gates in any universal set act on finitely
many qubits.

As for the error model, we will content ourselves with the simplest
scenario. Namely, some $\eta \in (0,1)$ is picked, and we assume that
each gate (including identity gates) undergoes an error with
probability $\eta$ and functions correctly with probability $1-\eta$. This
is known as the assumption of {\em local stochastic faults}
\cite{ab}. Quantum operations that describe errors within this
model have the form $\bT(\rho) = (1-\eta)\rho + \eta \bT'(\rho)$,
where $\bT'$ is a quantum operation whose Kraus operators
act nontrivially only on the qubits participating in a particular quantum gate.

Under this simple model, the occurrence of errors in space and time is
described naturally by a process known as {\em Bernoulli site
  percolation} \cite{hug} on the interaction graph $\Gamma_C$. Namely,
each vertex $v \in V_C$ is {\it occupied} with probability $\eta$ and
{\it vacant} with probability $1-\eta$, independently of all other
vertices, so that the occupied vertices correspond to the locations of
errors. Configurations of occupied and vacant vertices are given by
elements of the sample space $\Omega_C := \set{0,1}^{V_C}$, where $0$
($1$) indicates a vacant (occupied) vertex; the probability of an
event $F \subseteq \Omega_C$ will be denoted by $\P_\eta(F)$ [note
  the explicit parametrization of all probabilities by the vertex occupation
  density $\eta$]. An edge $v \lr v'$ is called {\it open} if both $v$ and
$v'$ are occupied, and {\it closed} otherwise. A set $K \subset V_C$
is called a {\it cluster} ({\it of connected occupied vertices}) if
all edges $v \lr v'$ with $v,v' \in K$ are open, and all edges $v \lr v'$
with $v \in K$, $v' \in V_C \bs K$ are closed. Among other things,
percolation theory is concerned with statistical properties of clusters as
the occupation density $\eta$ is varied. 

By construction, the interaction graph $\Gamma_C$ and the
corresponding site percolation process have the following properties:
(i) $\Gamma_C$ is {\it statistically homogeneous}, i.e., the
probability that a vertex is occupied does not depend on its location
in $\Gamma_C$, and (ii) the number of vertices that
can be reached from a given vertex by paths of length $n$ or less is
$O(2^{cn})$ for some $c < 1$. Then it follows from general arguments
of percolation theory \cite{hug} that: (1) there exists a number $\eta_* > 0$ (the {\it
  percolation threshold} \cite{note1}) such that, in the limit of an
infinite number of vertices, the
probability of a given vertex belonging to an infinite cluster of occupied
vertices is zero for $\eta < \eta_*$, and strictly greater than zero
for $\eta > \eta_*$; and (2) for $\eta < \eta_*$, the expected cluster
size is finite.

We now show that an estimate of the threshold error
rate may be derived by means of a renormalization argument common in
percolation theory (see, e.g., Chap.~4 of \cite{hug} for
details). The basic idea is the following: an (infinite) graph $\Gamma$ is
``coarse-grained'' by means of a rule that replaces suitably defined
groups of vertices of $\Gamma$ with single vertices, prescribes how
these new vertices are to be connected by edges, and defines a new
occupation probability $\eta'$ on the resulting graph $\Gamma'$.
Typically, if we start with a subcritical ($\eta < \eta_*$)
percolation proces on $\Gamma$, the net result of iterating the
renormalization process will be to push the system away from $\eta_*$
towards the ``trivial'' behavior ($\eta \rightarrow 0$).

Given a quantum circuit $C_0$, consider the single-layer encoding $C_1
= M(C_0)$ and the accompanying quantum computation code. Let $A$ be
the maximum number of locations involved in a single procedure in
$C_1$, $d$ the maximum number of errors corrected by the code, $s$
the spread of the code, and $k = \lfloor d/s \rfloor$.  Fix a
particular procedure in $C_1$. Then, if
more than $d$ qubits are in error at the end of this procedure, the
recovery stage of the subsequent procedure will fail. This will
happen precisely when at least $k+1$ errors occur during the
procedure. Denote this event by $E$. Using subadditivity of the
probability measure $\P_\eta(\cdot)$ and properties of the binomial
coefficients, we readily obtain the bound
$$
\P_\eta(E) \le
\sum^A_{\ell = k+1} {A \choose \ell} \eta^\ell (1-\eta)^{A-\ell} \le
2^A \eta^{k+1}.
$$
Then, for any $\eta$ satisfying the {\em threshold
  condition} $\eta < \eta_c \equiv 2^{-A/k}$,
we have $\P_\eta(E) < \eta$, i.e., the probability of $k$ or
more errors during a procedure is smaller than the probability of a
single error.

The influence of concatenation on the effective error rate can now be
understood as follows. Consider the interaction graph $\Gamma_{C_1}$
associated with the circuit $C_1 = M(C_0)$. We ``renormalize'' it
by replacing the vertices corresponding to each procedure in $C_1$
with a single vertex, and then by drawing edges appropriately. From
considerations of self-similarity we can expect that the resulting
renormalized graph $R(\Gamma_{C_1})$ is isomorphic to
$\Gamma_{C_0}$. We then say that a vertex $v$ of $R(\Gamma_{C_1})$ is
occupied if at least $k+1$ errors occur in the corresponding
procedure in $C_1$, and vacant otherwise. We will denote the occupation density
in $R(\Gamma_{C_1})$ by $R(\eta)$; it is related to the ``original''
occupation probability $\eta$ through $R(\eta) = \P_\eta(E)$.

When $\eta < \eta_c$, $R(\eta) \le 2^A \eta^{k+1} < \eta$.
That is, iteration of the renormalization process will keep reducing
the occupation probability further, as $R^{r+1}(\eta) < R^r(\eta) <
\ldots < \eta$. Better upper bounds on the renormalized occupation densities
can be easily computed by solving the recurrence $R^{r+1}(\eta) \le
2^A [R^r(\eta)]^{k+1}$ with the initial condition $R^0(\eta) = \eta$,
which yields
\begin{eqnarray*}
R^r(\eta) &\le& 2^{A [1 + (k+1)+\ldots + (k+1)^{r-1}]}\eta^{(k+1)^r}
\\
&=& 2^{A [(k+1)^r-1]/k} \eta^{(k+1)^r} \\
&=& 2^{-A/k} \left[2^{A/k}\eta \right]^{(k+1)^r}. 
\end{eqnarray*}

Note that the occurrence of errors on the coarse-grained interaction
graph is also modeled by a site percolation process because the errors
in any given procedure are assumed to occur independently of the
errors in all other procedures, and our renormalization transformation
has been defined only in terms of the restriction of global error
configurations to individual procedures. After $r$ levels of
concatenation, the effective error rate is exactly the site occupation
probability on $R^r(\Gamma_{C_0})$, and is bounded from above by
$c^{-1}_1(c_1\eta)^{(1+c_2)^r}$, where $c_1,c_2$ are constants related to the particular quantum
computation code used. The threshold condition on the error rate is
therefore that $\eta < c^{-1}_1$.

The necessary number of concatenation levels can now be determined in
the usual way \cite{preskill,ab}: if the unencoded
circuit $C_0$ has $N$ gates, then the encoded circuit $C_r$ will have
$N$ $r$-procedures, so that if we want the net computation error to be less
than some $\epsilon > 0$, the effective error rate must be at most $\epsilon / N$. This is guaranteed whenever $R^r(\eta)
< \epsilon / N$, which yields $r=\polylog (N/\epsilon)$
\cite{note2}. That is, the complexity of the
fault-tolerant quantum circuit will be larger than that of the
unencoded circuit by a polylogarithmic factor.

\begin{figure}
\includegraphics[width=0.9\columnwidth]{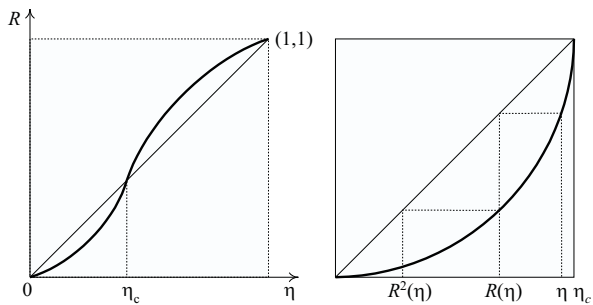}
\caption{Left: the ``S-shaped'' curve of the renormalization
  transformation $R$. Right: the portion of the curve for $0 \le\eta
  \le \eta_c$ and the ``staircase construction'' used to
  illustrate how the effective error rate is reduced
  under iteration of $R$.}
\label{fig:sshape}
\end{figure}

The above estimate of the threshold is quite
crude. The exact value of $\eta_c$ can be determined by finding nontrivial fixed
points of the renormalization transformation $\mapi{\eta}{R(\eta)}$
(note that the trivial values $\eta = 0,1$ are also fixed points of $R$). It is not hard to see that the graph of $R$ has the
``S-shape'' shown in Fig.~\ref{fig:sshape}. The exact value $\eta_c$ of the
threshold error rate is precisely the nontrivial fixed point of
$R$. One can also use the ``staircase construction'' pictured there to
see how the effective error rate goes down with the number of
concatenation levels.

Further information on the renormalized error rate
$R(\eta)$ can be obtained by means of general arguments of percolation theory. To this end let us observe the following properties of the event $E$: (i) its occurrence
depends on the status (i.e., occupied or vacant) of a finite number
$\alpha$ of vertices, and (ii) it is an {\it increasing event}, i.e., stable under
addition of more occupied vertices. Then the following
differential inequality holds for $0 < \eta < 1$ \cite{ms,ccfs}:
\begin{equation}
\frac{R(\eta) \left(1-R(\eta)\right)}{\eta(1-\eta)} \le R'(\eta) \le
\sqrt{\frac{\alpha}{\eta(1-\eta)}},
\label{eq:diffbounds}
\end{equation}
where the prime denotes differentiation with respect to
$\eta$. It follows easily from the first inequality in
(\ref{eq:diffbounds}) that $\lambda \equiv R'(\eta_c) \ge 1$; in fact,
a more careful analysis shows that $\lambda > 1$ \cite{gri}. From the second inequality in
(\ref{eq:diffbounds}) we get the following relation holds between
$\lambda$, $\alpha$, and $\eta_c$:
\begin{equation}
\lambda \le \sqrt{\frac{\alpha}{\eta_c (1-\eta_c)}}.
\label{eq:lambda}
\end{equation}

Consider now a quantum computer operating in the subthreshold regime, but
very close to the threshold, i.e., $\eta_c - \delta < \eta < \eta_c$
for some small $\delta > 0$. The minimum number of
concatenation levels necessary to bring the effective error rate down
to some suitable small value $\epsilon$ (where we can assume that
$\delta \ll \eta_c - \epsilon$) is controlled by
$\lambda$ (see the staircase construction in
Fig.~\ref{fig:sshape}). Linearizing $R$ around $\eta_c$ as $R(\eta) = \eta_c +
\lambda (\eta - \eta_c) + o(\abs{\eta-\eta_c})$ and iterating this process $r$
times, we see that we need to pick $r$ such that $\eta_c -\epsilon < \lambda^r
\delta$, i.e., it suffices to take
\begin{equation}
r \simeq \left\lfloor \frac{\log[(\eta_c - \epsilon)/\delta]}{\log\lambda}
\right\rfloor.
\label{eq:r}
\end{equation}
Solving (\ref{eq:r}) for $\lambda$, substituting into
(\ref{eq:lambda}), and using the fact that $(\eta_c - \epsilon)^{2/r}
\ge (\eta_c - \epsilon)^2$ for $r \ge 1$ and $0 < \epsilon < \eta_c < 1$,
we get
\begin{equation}
\eta_c (1-\eta_c) (\eta_c -\epsilon)^2 \le \alpha\delta^{2/r}.
\label{eq:tradeoff}
\end{equation}
This inequality can be regarded as a {\em threshold-overhead
  tradeoff inequality}: it shows that fault-tolerant quantum
  circuits with low concatenation overhead must have correspondingly small
  threshold error rates, and conversely that fault-tolerant quantum
  circuits with large threshold rates have the distinct disadvantage
  of requiring high concatenation overhead.

Let us close with a few remarks. First of all, we have so far paid no
attention to the percolation threshold $\eta_*$ on $\Gamma_C$ and how
it compares to $\eta_c$. We expect that
$\eta_c \ll \eta_*$ on the grounds that the (connected) clusters of
malfunctioning gates should be highly dilute in order for error
correction to succeed. Also, the methods used in this paper are
applicable to noisy classical computers as well. (This is,
  ultimately, not very surprising in light of the fact that the local
  stochastic error model is, essentially, classical in spirit.) It
would be interesting and useful to extend the methods used in this
paper (and percolation-theoretic methods in general) to less idealized
models of noise in quantum computers. A promising step in this
direction would be an application of percolation techniques to problems
involving graph-theoretic models of multiparticle entanglement
\cite{qgraphs1} and quantum computation \cite{qgraphs2}.

\paragraph*{Acknowledgements. ---} The author would like to thank
V.E.~Korepin for an invitation to speak at the Simons Conference. This
work was supported by the Defense Advanced Research Projects Agency
and by the U.S. Army Research Office.


\begin{thebibliography}{99}
\bibitem{note1}We are deviating from the common
    practice in percolation theory of using the subscript `$c$' to
    denote the percolation threshold because it is common practice in
    the theory of fault-tolerant quantum computation to use this
    subscript for the threshold error rate.
\bibitem{note2}The notation $x=\polylog(y)$ is used
  whenever we can find positive constants $k_1,k_2$, such that $x =
  k_1 (\log y)^{k_2}$.
\end{thebibliography}
\end{document}